# The Import and Export of *Cognitive Science*

Robert L. Goldstone

Indiana University

Department of Psychology, Indiana University, Bloomington, IN. 47405; rgoldsto@indiana.edu

Loet Leydesdorff

Amsterdam School of Communications Research (ASCoR), University of Amsterdam, Kloveniersburgwal 48, 1012 CX Amsterdam, The Netherlands; loet@leydesdorff.net; http://www.leydesdorff.net

* Corresponding author: rgoldsto@indiana.edu;    (812) 855-4853

Running head: The Import and Export of *Cognitive Science*

From its inception, a large part of the motivation for *Cognitive Science* has been the need for an interdisciplinary journal for the study of minds and intelligent systems. In the inaugural editorial for the journal, Allan Collins (1977) wrote "Current journals are fragmented along old disciplinary lines, so there is no common place for workers who approach these problems from different disciplines to talk to each other" (p. 1). The interdisciplinarity of the journal has served a valuable cross-fertilization function for those who read the journal to discover articles written for and by practitioners across a wide range of fields. The challenges of building and understanding intelligent systems are sufficiently large that they will most likely require the skills of psychologists, computer scientists, philosophers, educators, neuroscientists, and linguists collaborating and coordinating their efforts.

One threat to the interdisciplinarity of Cognitive Science, both the field and journal, is that it may become, or already be, too dominated by psychologists (Schunn, Crowley, & Okada, 1998; Von Eckardt, 2001). One piece of evidence supporting this contention is that many of the manuscripts submitted to *Cognitive Science* are given "psychology" as field keyword by their authors. In 2005, psychology was a keyword for 51% of submissions, followed distantly by linguistics (17%), artificial intelligence (13%), neuroscience (10%), computer science (9%), and philosophy (8%) (these percentages sum to more than 100% because authors are not restricted to designating only a single field).

Another quantitative way to assess the interdisciplinarity of *Cognitive Science* as well as its general intellectual niche is to analyze aggregated journal-journal citations. The Institute for Scientific Information (ISI) gathers data not only on how individual articles cite one another, but also on macroscopic citation patterns among journals. Journals or sets of journals can be considered as proxies for fields. As fields become established, they often create journals (Leydesdorff, Cozzens, & Van den Besselaar, 1994). As Collins (1977) wrote when launching *Cognitive Science*, "In starting the journal we are just adding another trapping in the formation of a new discipline" (p. 1). By studying the patterns of citations among journals that cite and are cited by *Cognitive Science*, we can better: 1) appreciate the scholarly ecology surrounding the journal and the journal's role within this ecology, 2) establish competitor and alternate journals, and 3) determine the natural clustering of fields related to cognitive science (Leydesdorff, 2006; forthcoming).

## Analyzing the Intellectual Ecology of *Cognitive Science*

The data we analyzed were the aggregated journal-journal citation records of the 7379 journals in the 2004 Journal Citation Report, including both the *Social Science Citation Index* and the *Science Citation Index*. The "Citing" data includes all of the journals (N=108) that were cited <u>by</u> articles appearing in *Cognitive Science* more than once in 2004.[1] It thus represents the intellectual <u>import</u> to *Cognitive Science*. The "Cited" data includes all of the journals (N=180) that in 2004 had more than one citation to articles appearing in *Cognitive Science*. The "Cited" data can be interpreted as the intellectual <u>export</u> of *Cognitive Science* to other journals. Square, asymmetrical matrices were created for both Cited and Citing data, with every cell containing the number of citations of one journal by the other. The matrices were analyzed using factor analysis with a varimax rotation in SPSS. The factors can be interpreted as the major clusters of journals related to *Cognitive Science*. We used the cosine instead of the Pearson correlation matrix for the visualization of citation patterns in Pajek (Salton & McGill, 1983; Ahlgren *et al.*, 2003). The cosine is convenient for the visualization because this measure has no negative values and normalizes on the geometrical instead of the arithmetical mean[2].

In terms of *Cognitive Science*'s import, in 2004 the journal cited 1090 articles appearing in 107 journals (one journal was dropped from the 108 journals cited by *Cognitive Science* because it was not cited by any other journal). The Citing data, plotted in Figure 1, reveal the main categories of journals that import references to *Cognitive Science*. A factor analysis revealed 5 prominent categories, and the impact of each of these categories is quantitatively measured by the proportion of variance in the citation matrix accounted for by each factor. The most prominent journal category is cognition (proportion of variance = 14.7%), followed by neuroscience and general science (13.4%), then social and general psychology (8.7%), then development and education (5.9%), and finally artificial intelligence (5.9%). The values for each of the journals cited by *Cognitive Science* on each of the five factors is graphically displayed at http://www.cogsci.rpi.edu/CSJarchive/Supplemental/index.html . We computed an import aggregate for each of these categories of journals by first classifying each journal according to the factor on which it has the greatest value, and then creating a cumulative count of the citations to these journal categories from *Cognitive Science* articles appearing in 2004 (see Cozzens & Leydesdorff, 1993 for the statistical construction of "macro-journals"). The resulting import is: experimental psychology (540 citations), neuroscience and general science (328), social and general psychology (90), development and education (57), and artificial intelligence (75). Figure 1 shows that *Cognitive Science* does not directly cite many education journals, but does cite journals that cite education journals. The computer science and cognitive psychology journals have relatively strong within-category citations and relatively weak cross-category citations.

---

[1] Single citations among journals are aggregated by the ISI under the category "All others".
[2] The cosine matrices based on the citation patterns of the ISI journal set are brought online at http://www.leydesdorff.net/jcr04/cited and http://www.leydesdorff.net/jcr04/citing, respectively. Pajek can be retrieved at http://vlado.fmf.uni-lj.si/pub/networks/pajek/ .

The Cited data, shown in Figure 2, indicates that *Cognitive Science* is cited by more journals than it cites. It has 1113 citations in 2004, spread across the 180 journals. A factor analysis yielded the following five factors in order of proportion variance accounted for: computer science/artificial intelligence (15.2%), cognition (7.6%), neuroscience and general science (5.0%), education (4.5%), and social/developmental psychology (3.8%). We computed an export aggregate analogously to how we calculated the import, creating a cumulative count of the 2004 citations to *Cognitive Science* for each category. The resulting export of citations is: computer science/artificial intelligence (429 citations), cognition (295), neuroscience and general science (130), education (172), and social/developmental psychology (87).

The two largest clusters of journals that cite *Cognitive Science* are cognitive psychology and computer science, with neuroscience rather densely interconnected with psychology. Education, linguistics, and philosophy journals all have stronger presences in the Cited data than the Citing data networks. Within the Cited network, *Cognitive Science* plays an important role in linking otherwise poorly related fields. This is seen upon visual inspection of Figure 2 and can be measured in terms of betweenness centrality, which measures the number of shortest paths connecting journals that include *Cognitive Science* as a node (Freeman, 1977). *Cognitive Science* is on the shortest paths between nodes of the network in 30.3% of the cases, whereas the second-largest betweeness value is 5.1% (for *Annual Review of Psychology*) and the average betweeness values for the other cosines among citation patterns in *Cognitive Science*'s citation network is 0.61% (Leydesdorff, in preparation). Thus, although *Cognitive Science* is cited highly in psychology, it has a strong export of citations to other fields, and it does satisfy at least one mission of an interdisciplinary journal – connecting together fields that might not otherwise efficiently exchange their knowledge.

To provide a preliminary exploration of the evolution of *Cognitive Science*'s export over time, we analyzed citations to *Cognitive Science* in the 1988 ISI database. A factor analysis revealed the following five factors in order of proportion variance accounted for: cognition (6.9%), artificial intelligence (6.0%), education (4.0%), developmental and social psychology (3.2%), and human-computer interaction (3.2%). *Cognitive Science* articles were cited 442 times in 1988, broken down by the five field factors as follows: cognition (101 citations), artificial intelligence (111), education (80), developmental and social psychology (56), and human-computer interaction (94).

## Discussion

From these data, we draw three conclusions about the role of *Cognitive Science* in cognitive science, and about the structure of cognitive science more broadly. First, within the journal ecology of *Cognitive Science*, the following fields robustly emerge: 1) computer science combined with artificial intelligence, 2) cognition, 3) neuroscience (combined with general science), 4) developmental/social/educational psychology (with developmental psychology sometimes cohering more tightly with social psychology, and sometimes with education). Although these fields reliably fall out of the factor analyses, a closer examination reveals that human-computer interaction played a larger role in *Cognitive Science* in 1988 than 2004. Conversely, neuroscience has a considerably greater influence on *Cognitive Science* in 2004 than it did in 1988. Journals related to language show up in several different categories within *Cognitive Science*'s journal ecology. Philosophy and linguistics, although they are core disciplines to cognitive science, have relatively small roles to play in the journal ecology of *Cognitive Science*. This is, in part, because not all philosophy and linguistics journals are included in ISI's general or social science indexes. However, *Cognitive Science* is more weakly connected to linguistics and philosophy journal that are indexed than it is to computer science, psychology or neuroscience journals.

Second, the specifics of the claim for the interdisciplinary relevance of *Cognitive Science* depend upon whether we refer to export or import. *Cognitive Science* is heavily cited by computer science and artificial intelligence journals, but does not cite these journals as much. In terms of import to *Cognitive Science*, the four most influential journal categories all relate to psychology. Note that the asymmetry in citations to and from computer science journals cannot be explained by computer science articles having fewer references than social science articles; *Cognitive Science* was cited 429 times by computer science articles, but only cited 75 computer science articles. Asymmetries in import and export factors are common in science (Boyack, Klavans, & Börner, 2005), and are not necessarily ominous signs (e.g. business management journals cite psychology journals far more than vice versa). *Cognitive Science's* import profile can be interpreted as suggesting that the journal is at some risk of being psychology-centric in the literature its articles use to build their arguments. Its export profile, however, suggests that the journal functions as a window for computer scientists on new developments in psychology and relevant fields in the social sciences.

Third, the claim for interdisciplinary importance is far stronger when considering betweeness. If *Cognitive Science* did not exist, the citation path connecting journals would be far longer. Admittedly, this analysis is predicated upon selecting journals that cite, or are cited by, *Cognitive Science*. Still, *Cognitive Science* plays a unique bridging role in efficiently transferring information across psychology, computer science, neuroscience, and education. In addition to gauging interdisciplinarity by measuring the number of articles with a multidisciplinary team (Schunn et al., 1998) or the number of disciplines that actively engage in cognitive science research (Von Eckardt, 2001), the betweeness measure offers yet another perspective on what it means for cognitive science to be interdisciplinary -- to play a critical role in connecting fields. Similar to how social networks benefit greatly by having individuals span and therefore connect otherwise disjoined cliques (Granovetter, 1973), the scientific network benefits from having intellectual communities that effectively merge perspectives, tools, and methods.

More generally, we are optimistic about the possibilities for applying cognitive science and scientometric analyses to understanding the structure and evolution of cognitive science itself. Even more broadly speaking, science is a web of interrelated fields, and interdisciplinary research stitches together scientific communities that are constantly under pressure of increasing fractionation. Counteracting the trend toward ever increasing specialization in science, interdisciplinary approaches promote meaning-building by creating rich, interconnected networks of knowledge. In much the same way that a concept or belief does not mean anything in isolation but rather only within a network of other concepts and beliefs (Quine & Ullian, 1970; Stich, 1983), the meaning of a scientific result is best gauged by measuring its specific impact on the entire knowledge network. Our techniques reveal these local citation impacts and to understand the mechanisms. Interdisciplinary fields such as cognitive science have wisely placed their bets on creating meaning not only by decomposing knowledge into smaller and smaller nuggets, but also by interrelating and connecting these nuggets.

**Author Notes**

We thank Robin Kramer for extensive work on preparing the figures and tables, and Katy Börner for helpful comments.  This research was funded by National Science Foundation grant 052792, and Department of Education, Institute of Education Sciences grant R305H050116.  Correspondence concerning this article should be addressed to rgoldsto@indiana.edu or Robert Goldstone, Psychology Department, Indiana University, Bloomington, Indiana 47405.  Further information about the laboratory of the first author can be found at http://cognitrn.psych.indiana.edu/.  For further information about the supporting software, visit http://www.leydesdorff.net/jcr04, and for further reading about these techniques, visit http://www.leydesdorff.net/list.htm.

Table 1: Journals that are cited by *Cognitive Science*, ordered by their factor of maximum value.

| Journal Number | Journal Name | Journal Number | Journal Name | Journal Number | Journal Name |
|---|---|---|---|---|---|
| 1 | Cognitive Science | 26 | Journal of Cognitive Neuroscience | 51 | Psychological Bulletin |
| 2 | Psychonomic Bulletin & Review | 27 | Current Biology | 52 | Annual Review of Psychology |
| 3 | Quarterly Journal of Experimental Psychology Section A - Human Experimental Psychology | 28 | Proceedings of the National Academy of Sciences | 53 | Organizational Behavior and Human Decision Processes |
| 4 | Cognitive Psychology | 29 | Cognitive Brain Research | 54 | Psychological Science |
| 5 | Journal of Experimental Psychology - General | 30 | Science | 55 | Current Directions in Psychological Science |
| 6 | Memory & Cognition | 31 | Philosophical Transactions of the Royal Society of London Series B - Biological Sciences | 56 | American Psychologist |
| 7 | Journal of Experimental Psychology - Learning Memory and Cognition | 32 | Nature | 57 | Human-Computer Interaction |
| 8 | Psychology of Learning and Motivation - Advances in Research and Theory | 33 | Journal of Neurophysiology | 58 | Cognitive Development |
| 9 | Journal of Memory and Language | 34 | Human Brain Mapping | 59 | Child Development |
| 10 | Psychological Review | 35 | Neural Computation | 60 | Developmental Psychology |
| 11 | Acta Psychologica | 36 | Visual Neuroscience | 61 | Developmental Science |
| 12 | Journal of Experimental Psychology - Human Perception and Performance | 37 | Neuroimage | 62 | Journal of Experimental Child Psychology |
| 13 | Visual Cognition | 38 | Journal of Physiology - London | 63 | Journal of Educational Psychology |
| 14 | Perception & Psychophysics | 39 | Brain | 64 | Journal of the Optical Society of America A - Optics Image Science and Vision |
| 15 | Cognition | 40 | Neuropsychologia | 65 | Journal of the Acoustical Society of America |
| 16 | Brain and Language | 41 | Behavioral and Brain Sciences | 66 | Computational Intelligence |
| 17 | Cognitive Neuropsychology | 42 | Experimental Brain Research | 67 | Artificial Intelligence |
| 18 | Nature Reviews Neuroscience | 43 | Vision Research | 68 | IEEE Transactions on Evolutionary Computation |
| 19 | Annual Review of Neuroscience | 44 | Investigative Ophthalmology & Visual Science | 69 | Machine Learning |
| 20 | Nature Neuroscience | 45 | IEEE Transactions on Biomedical Engineering | 70 | Computational Linguistics |
| 21 | Neuroreport | 46 | Neurology | 71 | IEEE Transactions on Pattern Analysis and Machine Intelligence |
| 22 | Neuron | 47 | Journal of the Learning Sciences | 72 | Journal of Experimental Psychology - Animal Behavior Processes |
| 23 | Cerebral Cortex | 48 | Journal of Personality and Social Psychology | 73 | Journal of the Experimental Analysis of Behavior |
| 24 | Journal of Neuroscience | 49 | Journal of Experimental Social Psychology | | |
| 25 | Trends in Cognitive Sciences | 50 | Personality and Social Psychology Bulletin | | |

Table 2: Journals Citing *Cognitive Science*, ordered by their factor of maximum value.

| Journal Number | Journal Name | Journal Number | Journal Name | Journal Number | Journal Name |
|---|---|---|---|---|---|
| 1 | Cognitive Science | 25 | Journal of Experimental Psychology - Human Perception and Performance | 49 | Cognitive Neuropsychology |
| 2 | IEEE Intelligent Systems | 26 | Memory | 50 | Journal of Neurophysiology |
| 3 | Autonomous Agents and Multi-Agent Systems | 27 | Acta Psychologica | 51 | Annals of the New York Academy of Sciences |
| 4 | Journal of Artificial Intelligence Research | 28 | Cognition | 52 | Nature |
| 5 | Artificial Intelligence | 29 | Visual Cognition | 53 | Journal of Motor Behavior |
| 6 | Computational Intelligence | 30 | Behavioral and Brain Sciences | 54 | Vision Research |
| 7 | IEEE Transactions on Software Engineering | 31 | Psychology and Aging | 55 | Proceedings of the Royal Society of London Series B - Biological Sciences |
| 8 | User Modeling and User-Adapted Interaction | 32 | Consciousness and Cognition | 56 | Hearing Research |
| 9 | IEEE Transactions on Neural Networks | 33 | Journal of Experimental Psychology - Applied | 57 | Journal of Applied Psychology |
| 10 | Computational Linguistics | 34 | Academic Medicine | 58 | Review of Educational Research |
| 11 | Neural Networks | 35 | Medical Education | 59 | Educational Psychologist |
| 12 | Neural Computation | 36 | Journal of Sport & Exercise Psychology | 60 | Journal of Educational Psychology |
| 13 | Adaptive Behavior | 37 | Neuropsychologia | 61 | Learning and Instruction |
| 14 | Annual Review of Information Science and Technology | 38 | Journal of Cognitive Neuroscience | 62 | Journal of the Learning Sciences |
| 15 | Journal of Experimental Psychology - General | 39 | Brain | 63 | Leadership Quarterly |
| 16 | Memory & Cognition | 40 | Cognitive Brain Research | 64 | Cognitive Development |
| 17 | Psychonomic Bulletin & Review | 41 | Neuroreport | 65 | Developmental Psychology |
| 18 | Journal of Experimental Psychology - Learning Memory and Cognition | 42 | Trends in Cognitive Sciences | 66 | Child Development |
| 19 | Cognitive Psychology | 43 | Cortex | 67 | Developmental Science |
| 20 | Quarterly Journal of Experimental Psychology Section A - Human Experimental Psychology | 44 | Trends in Neurosciences | 68 | Journal of Experimental Child Psychology |
| 21 | Psychological Review | 45 | Neuroimage | 69 | Journal of Child Psychology and Psychiatry and Allied Disciplines |
| 22 | Psychology of Learning and Motivation - Advances in Research and Theory | 46 | Brain and Language | 70 | Annual Review of Psychology |
| 23 | Journal of Memory and Language | 47 | Journal of Neurolinguistics | 71 | Journal of Personality and Social Psychology |
| 24 | Psychological Science | 48 | Experimental Brain Research | 72 | Advances in Experimental Social Psychology |

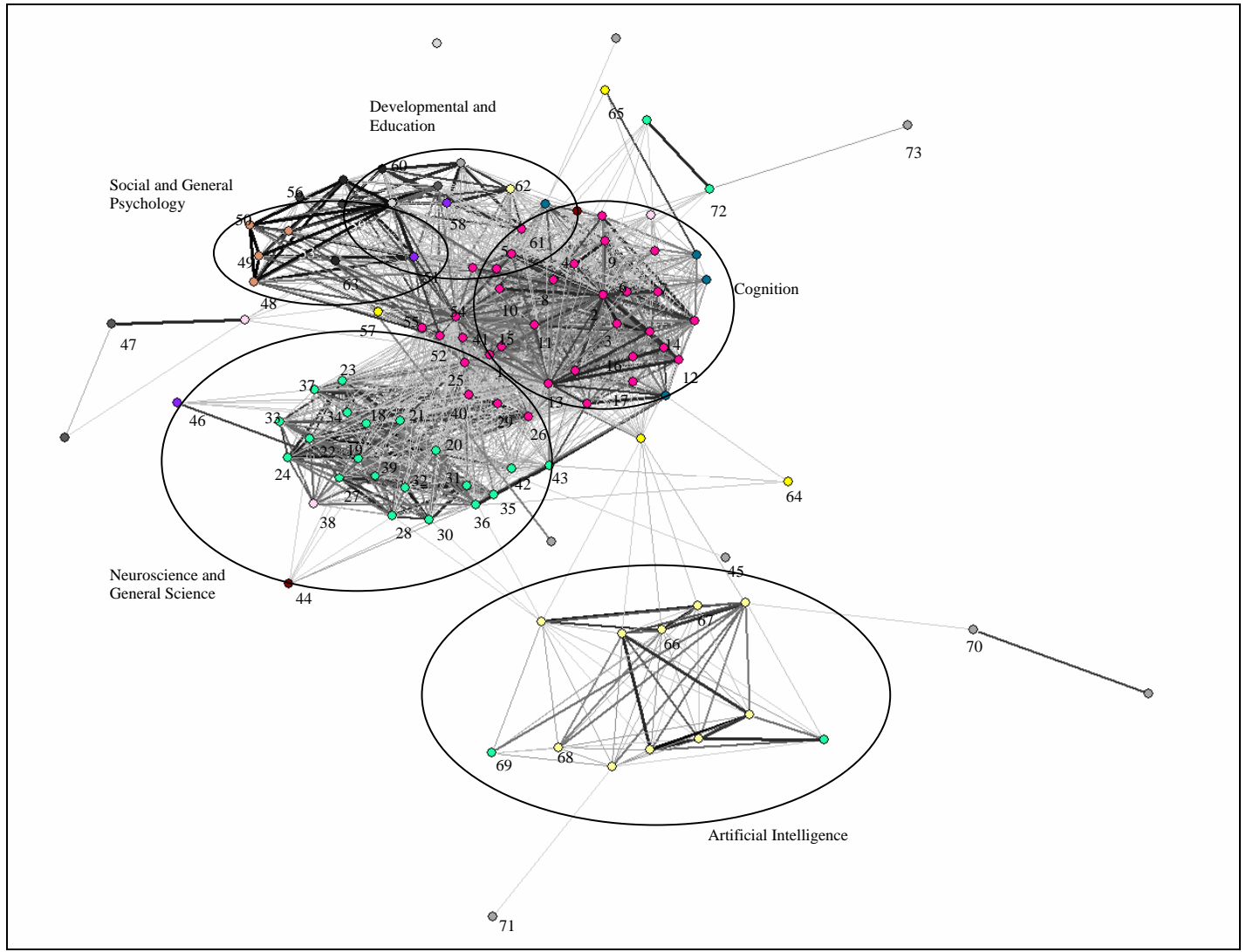

**Figure 1**: The network of 107 journals cited by *Cognitive Science* in 2004. The colors/shades of the nodes are provided by Pajek's graph-based clustering algorithm. The ellipses show clusters obtained by a factor analysis on the journal-to-journal citation export matrix. Edges are only shown between journals that have a citation pattern with a cosine greater than 0.2, and the thickness of an edge reflects the citation proximity/cosine of the connected journals. The specific journals can be identified by looking up their numbers in Table 1. Only journals with ISI Impact Factors greater than 1.4 are assigned numbers.

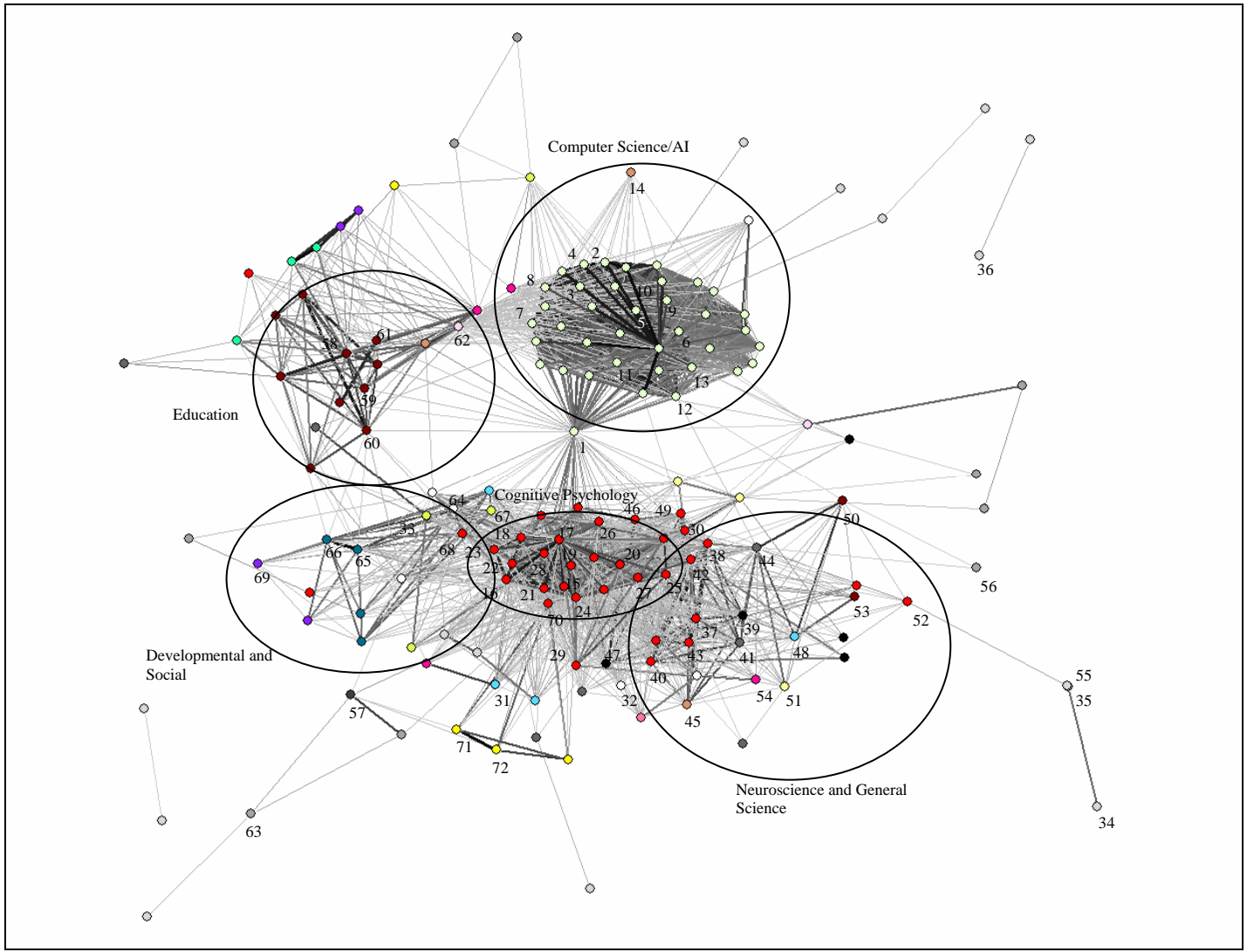

**Figure 2**: The network of 164 journals citing *Cognitive Science* in 2004. As with Figure 1, edges connecting journals reflect strong cross-citation patterns. Journals with Impact Factors greater than 1.4 can be identified by looking up their numbers in Table 2.